\documentclass[11pt]{article}
\usepackage{axodraw}
\usepackage{epsfig}
\usepackage{amsfonts}
\usepackage{amsmath}
\usepackage{bbm}
\usepackage{cite}
\allowdisplaybreaks[1]

\input pix.sty

\newcommand{\ds}{d_s}
\newcommand{\T}{\rmii{$T$}}

\renewcommand{\eq}{eq.~}
\renewcommand{\eqs}{eqs.~}

\renewcommand{\se}{sec.~}

\renewcommand{\fig}{fig.~}

\newcommand{\Nc}{N_{\rm c}}

\newcommand{\gammaE}{\gamma_\rmii{E}}

\newcommand{\rmO}{{\mathcal{O}}}
\newcommand{\bmu}{\bar\mu}

\newcommand{\CF}{C_\rmii{F}}

\def\lsi{\raise0.3ex\hbox{$<$\kern-0.75em\raise-1.1ex\hbox{$\sim$}}}
\def\gsi{\raise0.3ex\hbox{$>$\kern-0.75em\raise-1.1ex\hbox{$\sim$}}}
\newcommand{\lsim}{\mathop{\lsi}}
\newcommand{\gsim}{\mathop{\gsi}}

\newcommand{\rmii}[1]{{\mbox{\tiny\rm{#1}}}}

\newcommand{\re}{\mathop{\mbox{Re}}}

\newcommand{\Tint}[1]{{\hbox{$\sum$}\!\!\!\!\!\!\!\int\,}_{\!\!\!\!\raise-0.9ex\hbox{$\scriptstyle{#1}$}}}
\newcommand{\Tinti}[1]{{{\Sigma}\!\!\!\!\raise0.3ex\hbox{$\int$}_\rmii{${#1}$}}}

\newcommand{\bi}{\begin{itemize}}
\newcommand{\ei}{\end{itemize}}
\newcommand{\hide}[1]{ }

\newcommand{\deltabar}{\delta\!\!\!\raise0.7ex\hbox{--}\,}
\def\TAsc(#1,#2)(#3,#4,#5)%
{\SetWidth{2.0}\CArc(#1,#2)(#3,#4,#5)\SetWidth{1.0}}
\def\Lwidth{3}

\def\TAgl(#1,#2)(#3,#4,#5){\SetWidth{2.0}\PhotonArc(#1,#2)(#3,#4,#5){\Lwidth}%
{6.283 #3 mul 360 div #4 #5 sub #4 #5 sub mul sqrt mul Tdensity mul}%
\SetWidth{1.0}}
\def\TLgl(#1,#2)(#3,#4){\SetWidth{2.0}\Photon(#1,#2)(#3,#4){\Lwidth}
{#1 #3 sub #1 #3 sub mul #2 #4 sub #2 #4 sub mul add sqrt Tdensity mul}%
\SetWidth{1.0}}

\renewcommand{\picc}[1]{\;\parbox[c]{60pt}{\begin{picture}(60,30)(0,-3)
\SetWidth{1.0}\SetScale{1.1} #1 \end{picture}}\; }
\def\Lwidth{1.3}

\makeatletter \@addtoreset{equation}{section} \makeatother

\makeatletter
\renewcommand\section{\@startsection {section}{1}{\z@}%
                                   {-5.5ex \@plus -1ex \@minus -.2ex}
                                   {2.3ex \@plus.2ex}%
                                   {\normalfont\large\bfseries}}
\renewcommand\subsection{\@startsection{subsection}{2}{\z@}%
                                     {-3.25ex\@plus -1ex \@minus -.2ex}%
                                     {1.5ex \@plus .2ex}%
                                     {\normalfont\normalsize\bfseries}}
\renewcommand\thesection {\@arabic\c@section}
\renewcommand\thesubsection   {\thesection.\@arabic\c@subsection}
\renewcommand{\@seccntformat}[1]{%
\csname the#1\endcsname.\hspace{1.0em}}
\makeatother

\begin{document}

\flushbottom

\begin{titlepage}

\begin{flushright}
October 2019
\end{flushright} 
\begin{centering}

\vfill

{\Large{\bf
 Non-relativistic susceptibility and a dark matter application
}} 

\vspace{0.8cm}

S.~Biondini$^{\rm a}$, 
Seyong Kim$^{\rm b,c}$ and 
M.~Laine$^{\rm c}$ 

\vspace{0.8cm}

$^\rmi{a}$%
{\em
Van Swinderen Institute, University of Groningen, \\
Nijenborgh 4, NL-9747 AG Groningen, the Netherlands\\}

\vspace*{0.3cm}

$^\rmi{b}$%
{\em
Department of Physics, 
Sejong University, Seoul 143-747, South Korea\\}

\vspace*{0.3cm}

$^\rmi{c}$%
{\em
AEC, Institute for Theoretical Physics, 
University of Bern, \\ 
Sidlerstrasse 5, CH-3012 Bern, Switzerland\\} 

\vspace*{0.8cm}

\mbox{\bf Abstract}

\end{centering}

\vspace*{0.3cm}
 
\noindent
When thermal rate equations are derived for 
the evolution of slow variables, it is often practical 
to parametrize the right-hand side with chemical potentials. 
To close the system, the chemical potentials are subsequently 
re-expressed in terms of the slow variables, which involves
the consideration of a ``susceptibility''. 
Here we study a non-relativistic situation in which 
chemical potentials are large compared with the temperature, as is
relevant for late-time pair annihilations in dark matter freeze-out. 
An order-of-magnitude estimate and a lattice simulation
are presented for a susceptibility dominated by bound states 
of stop-like mediators. After this ``calibration'', the formalism 
is applied to a model with Majorana singlet dark matter, confirming 
that masses up to the multi-TeV domain are viable in the presence of 
sufficient (though not beyond a limit) 
mass degeneracy in the dark sector.

\vfill

%
 

\vfill

\end{titlepage}

%
\section{Introduction}

In the weakly interacting massive particle (WIMP) scenario, 
the number density of the dark sector
is usually assumed to satisfy 
the so-called Lee-Weinberg equation~\cite{clas1}, 
\be
 \dot{n} + 3 H n = - \langle \sigma v \rangle 
 \, \bigl( n^2 - n_\rmi{eq}^2 \bigr) 
 \;, \la{lw}
\ee
where $H$ is the Hubble rate. Eq.~\nr{lw} can be derived from
Boltzmann equations, assuming kinetic equilibrium and 
integrating over momenta~\cite{clas2,old1}. 
However, Boltzmann equations 
have a limited range of validity, failing e.g.\ 
if interactions within the dark sector become strong. 

If the interactions are strong enough 
to form bound states, a standard practice is
to add bound states as additional degrees of freedom in a set
of Boltzmann equations~\cite{old32,old4}. 
However, there are challenges with this approach.  
One problem is that strongly interacting systems 
have {\it many} bound states; another is that their number
varies with the temperature~\cite{seq}; a further one is that 
bound-state rate coefficients are cumbersome to compute. 
It would be attractive to have 
a more ``inclusive'' framework which does not require 
{\it a priori} knowledge of how many (if any) bound states are 
present, even if at very low temperatures a set of coupled 
equations surely becomes necessary.  

One way to promote \eq\nr{lw} beyond Boltzmann
equations is to note that the coefficient $\langle \sigma v \rangle$
is independent of the value of the dynamical variable $n$.  
Thus, one can assume that the system is prepared in a state 
close to equilibrium, and linearize in deviations. Thereby we can make
contact with linear response theory, which permits to 
define a chemical equilibration rate, 
$
 \Gamma^{ }_\rmi{chem} = 
 2 n^{ }_\rmi{eq}\langle \sigma v \rangle
$, 
on a non-perturbative level~\cite{chemical}. 
Furthermore, within the 
non-relativistic expansion~\cite{bodwin}, $\Gamma^{ }_\rmi{chem}$ can be 
related to the thermal expectation value of a local annihilation
operator, and then be measured with lattice simulations
if necessary~\cite{swave}. 

Another generalization of \eq\nr{lw} was put forward in ref.~\cite{binder}.
Making use of Schwinger-Keldysh formalism, 
which goes beyond linear response theory,  
the authors reproduced the expression of 
ref.~\cite{swave} for $\langle \sigma v \rangle$, 
but in addition suggested that the functional form should read
\be
 \dot{n} + 3 H n = - \langle \sigma v \rangle 
 \bigl(  e^{2\beta\mu(n)}  - 1 \bigr) n^{2}_\rmi{eq}
 \;, \la{comb}
\ee
where $\beta \equiv 1/T$ and $\mu$ couples 
to the total number of dark sector 
particles. In a weakly coupled system, 
$ 
 e^{\beta\mu} n^{ }_\rmi{eq} \approx n
$ (cf.\ \eq\nr{n_actual}), 
but in general this need not 
be the case. The relation between 
$n$ and $\mu$ leads to a variant of the Saha equation, 
familiar from the physics of recombination,
displaying significant modifications if $T \lsim \Delta E$, 
where $\Delta E$ is a binding energy. 

In general, 
the quantity $\partial n / \partial \mu$ is called a 
``susceptibility''.  
In many cosmological problems, such as leptogenesis, we find ourselves in 
the regime $\mu^{ } \ll T$; susceptibilities 
for this situation have been worked out up to higher 
perturbative orders~\cite{kubo,sangel}. 
For WIMPs, it is the non-relativistic regime $\mu \sim M \gg T$
that needs to be attacked. The goal of the present study is
to define and estimate a susceptibility for the latter
situation, and to show how the corresponding 
result can be implemented in a dark matter computation
employing \eq\nr{comb}.

%
\section{General setup}

We consider a theory whose dark sector contains a {\em charged field},
whose quanta may be called particles and antiparticles. This charged
field plays the role of a ``mediator'', i.e.\ it couples dark matter
to Standard Model particles.  If the coupling goes through 
Yukawa interactions, the mediator has the charge assignment of
one of the Standard Model fields, for instance that of a right-handed
top quark. We assume that the mediator interacts strongly through
SU($\Nc^{ }$) gauge theory. Its gauge coupling is denoted by $g^2$,
the Casimir coefficient of 
the fundamental representation by $\CF^{ }\equiv(\Nc^2 -1)/(2\Nc^{ })$, 
and we let $\alpha \equiv g^2 \CF^{ }/(4\pi)$. 
 
Let $\hat{\theta}$ and $\hat{\eta}$ be field
operators which annihilate particles 
and antiparticles of the charged field, 
respectively, and define the number density
operator by 
\be
 \hat{N} = \int_\vc{x} \hat{n}(\vc{x})
 \;, \quad
 \hat{n}(\vc{x}) \; \equiv \; \hat{\theta}^\dagger\hat{\theta} 
 + \hat{\eta}^\dagger\hat{\eta} 
 \;. \la{N_def}
\ee
Moreover we denote by $n^{ }_\rmi{eq}$ 
the expectation value of $\hat{n}$
in full chemical equilibrium, i.e.
\be 
 n^{ }_\rmi{eq} \; \equiv \; \lim_{\mu\to 0} 
 \langle \hat{n} \rangle
 \;. \la{neq_def}
\ee
The role of $\mu$ is defined through \eq\nr{rho_def}. 
We assume that 
$\theta$ and $\eta$ have $\ds^{ }\Nc^{ }$ real components
($\Nc^{ }\equiv 3$), 
where $\ds^{ } \equiv 2s+1$  
is the degeneracy of spin degrees of freedom.

Because the processes which change the number density
are very slow,\footnote{%
 By slow we mean slow compared with processes responsible for kinetic 
 equilibration, and with reactions between Standard Model particles; that is, 
 the number density is assumed to be the only non-equilibrium variable. 
 }
it is appropriate to consider a state of
the system in which $n \neq n^{ }_\rmi{eq}$. This can be 
imposed by coupling $\hat{N}$ to a chemical potential, 
so that the density matrix has the form 
\be
 \hat{\rho} \; \equiv \; \frac{ \exp[-\beta (\hat{H} - \mu\hat{N})] }{Z} 
 \;, 
 \la{rho_def}
\ee
where the partition function is given by 
$
 Z = \tr e^{-\beta (\hat{H} - \mu\hat{N})}
$.
In the thermodynamic limit the partition function can be parametrized
by the pressure $p$ as 
$
 Z = e^{p \beta V}
$,
where $V$ is the spatial volume. 
The number density is obtained as 
\be
 n(\mu)
   \; = \; \frac{\partial p}{\partial \mu} 
   \; = \; \frac{\langle \hat{N} \rangle}{V}
   \; = \; \bigl\langle \hat{n}(\vc{0}) \bigr\rangle
   \;, \quad
   \langle ... \rangle 
   \; \equiv \; 
   \tr[\hat{\rho}(...)]
   \;, \la{n}
\ee
where we assumed the system to be translationally invariant. 
A susceptibility is defined as 
\be
 \chi \; \equiv \; T\, \frac{\partial n}{\partial\mu} 
 \; = \;  
  \frac{\langle \hat{N}^2 \rangle - \langle \hat{N} \rangle^2}{V}
 \; = \;  
 \int_\vc{x} 
 \Bigl\{ 
 \bigl\langle \hat{n}(\vc{x})\, \hat{n}(\vc{0}) \bigr\rangle
 \; - \; 
 \bigl\langle \hat{n}(\vc{0}) \bigr\rangle^2_{ }
 \Bigr\}
 \;. \la{chi}
\ee

We now formally expand the pressure in a fugacity expansion, 
\be
 p = p^{ }_0 + p^{ }_1 \, e^{\beta\mu} + p^{ }_2 \, e^{2\beta\mu} + \ldots
 \;,  
\ee
where $p^{ }_n \sim e^{-n M/T}$ (cf.\ \eq\nr{n_canonical})
and $M$ is the dark matter mass scale. 
Let us assume that the coefficient $p^{ }_2$ could be anomalously large
because of a bound-state contribution. 
{}From \eqs\nr{n}, \nr{chi}, 
the corresponding expansions 
for $n$ and $\chi$ read
\ba
 n T & = &  p^{ }_1 \, e^{\beta\mu} + 2 p^{ }_2 \, e^{2\beta\mu} + \ldots 
 \;, \la{n_exp} \\
 \chi T & = &  p^{ }_1 \, e^{\beta\mu} + 4 p^{ }_2 \, e^{2\beta\mu} + \ldots
 \;. \la{chi_exp}
\ea
{}From \eq\nr{n_exp}, omitting $p^{ }_3$ and higher-order terms,\footnote{%
 This omission, also made in ref.~\cite{binder}, corresponds to the assumption
 that $n$-body bound states of the heavy particles (here $n \ge 3$) 
 have binding energies much smaller than $M$, so that they carry 
 a minor fraction of the total dark matter number density and do
 not substantially contribute to the pair annihilation process. This
 should be well justified for $M \sim $~TeV~$\gg$~GeV.
 }
we get, 
in accordance with ref.~\cite{binder}, 
\be
 e^{\beta\mu} \approx \frac{-p^{ }_1 
 + \sqrt{p_1^2 + 8 p^{ }_2 n T}}{4 p^{ }_2}
 \;.
\ee
Moreover, by subtracting \eq\nr{chi_exp} from \nr{n_exp},
we can estimate the coefficient $p^{ }_2$ as
\be
 2 p^{ }_2 \, e^{2\beta\mu} \approx T (\chi - n)
 \;. \la{p2}
\ee

In practical applications, it is convenient to 
remove exponentially small terms by noting that 
in the limit of chemical equilibrium, 
when effects suppressed by $e^{-M/T}$ can be omitted, we can 
identify
$ 
 p^{ }_1 \; = \; n^{ }_\rmi{eq} T
$
(cf.\ \eq\nr{n_exp}).
Moreover we can define
\be
 p^{ }_2 \; \equiv \; \hat{p}^{ }_2 \, n^{2}_\rmi{eq} T
 \;. \la{p2_hat}
\ee 
Then the combination appearing in \eq\nr{comb} becomes
\be
 e^{\beta\mu} n^{ }_\rmi{eq} \; \approx \;  
 \frac{ - 1 + \sqrt{1 + 8 \hat{p}^{ }_2 n} }
 {4 \hat{p}^{ }_2} 
 \; = \; 
 \frac{2 n }{
   1 + \sqrt{ 1 + 8 \hat{p}^{ }_2 n}
 } 
 \;. \la{n_actual}
\ee
In perturbation theory, $\hat{p}^{ }_2$ is generated by interactions.
In a weakly coupled system, we may expect it 
to be small, in which limit \eq\nr{n_actual} reduces to 
$
 e^{\beta\mu} n^{ }_\rmi{eq} \approx n
$. 

%
\section{Order-of-magnitude estimate}
\la{se:pert}

In order to estimate the magnitude of $\hat{p}^{ }_2$, 
it is useful to employ the canonical formalism. 
Let us denote the eigenstates of the Hamiltonian by 
\be
 | \, n^{ }_\theta, n^{ }_\eta \, \rangle
 \;, \quad 
 n^{ }_\theta,n^{ }_\eta \in \{ 0,1,2,... \}
 \;, 
\ee
where $n^{ }_\theta,n^{ }_\eta$ enumerate the $\theta$ and $\eta$
particles present. 
We assume that in a dilute system 
($T \ll M$) the observables $n,\chi$ are dominated by three sectors
of the Fock space, namely 
$(n^{ }_\theta ,n^{ }_\eta) = (1,0), (0,1), (1,1)$, 
whereas the contributions of the sectors 
$(n^{ }_\theta ,n^{ }_\eta) = (2,0), (0,2)$ and those of any
three-particle and higher states are Boltzmann-suppressed.
As this simplifies formal manipulations, 
we stay in a finite volume for a moment, 
so that one-particle states are parametrized 
by a set of discrete momenta $\{\vc{p}^{ }_\theta \}$, with 
the corresponding degeneracies 
$c^{ }_\theta = c^{ }_\eta  = \ds^{ }\Nc^{ }$. 
The two-particle
states can be either bound or scattering states; for brevity, we
use a scattering-like notation here, parametrizing the states with 
a pair of momenta $(\vc{p}^{ }_\theta,\vc{p}^{ }_\eta)$, and denoting
by $E^{ }_{p_\theta,p_\eta}$ the corresponding energy and by
$c^{ }_{\theta,\eta}$ the degeneracy factor. With this notation, 
and sticking to a state normalization without volume factors
in order to avoid clutter, 
the number density of \eq\nr{n} can schematically be evaluated as 
\be
 n \;\simeq\; \frac{1}{V}
 \frac{ 
   \sum_{\vc{p}^{ }_\theta} 
   c^{ }_\theta\, 
   e^{\beta (\mu - E^{ }_{p_\theta} ) } 
 + 
   \sum_{\vc{p}^{ }_\eta} 
   c^{ }_\eta\, 
   e^{\beta (\mu - E^{ }_{p_\eta} ) } 
 + 
  2 \sum_{\vc{p}^{ }_\theta,\vc{p}^{ }_\eta}
   c^{ }_{\theta,\eta}\,
   e^{\beta (2\mu - E^{ }_{p_\theta,p_\eta} ) }  
 }{
 1
 +  
   \sum_{\vc{p}^{ }_\theta} 
   c^{ }_\theta\, 
   e^{\beta (\mu - E^{ }_{p_\theta} ) } 
 + 
   \sum_{\vc{p}^{ }_\eta} 
   c^{ }_\eta\, 
   e^{\beta (\mu - E^{ }_{p_\eta} ) } 
 }
 \;. \la{n_canonical}
\ee
The factor $2$ in the third term of the numerator 
emerges because there are two
particles in the sector $n^{ }_\theta = n^{ }_\eta = 1$.
The denominator represents normalization 
by $Z$ (cf.\ \eq\nr{rho_def}); 
the first term originates from 
the sector $n^{ }_\theta = n^{ }_\eta = 0$. 
Since we need to go up to 
second order in the fugacity expansion, we need to include 
the next terms as well.  
Eq.~\nr{n_canonical} represents a relation between $n$ and $\mu$, 
and is as such a variant of the Saha equation, 
even if the Saha equation
is usually used in a different way.\footnote{%
 Normally one eliminates 
 $e^{\beta\mu}$ in favour of the number densities of unbound states
 ($n^{ }_\theta$ and $n^{ }_\eta$), 
 {\it viz.}\ 
 $
  e^{\beta\mu} = \frac{n^{ }_\theta}{c^{ }_\theta}
  \bigl( \frac{2\pi}{M^{ }_\theta T} \bigr)^{\fr32}
  e^{\beta M^{ }_\theta}
 $. 
 Then the last term in the numerator of \eq\nr{n_canonical} 
 is proportional to $n^{ }_\theta n^{ }_\eta e^{\beta \Delta E}$, 
 where $\Delta E$ is the binding energy and we assume the
 existence of {\em one} bound state. Subsequently, if the 
 relation of  $n^{ }_\theta$ and  $n^{ }_\eta$ is known
 (in our case $n^{ }_\theta = n^{ }_\eta$), they 
 can be solved for as a function of the total number density $n$
 and the exponential factor $e^{\beta \Delta E}$. 
 Here we instead want to solve for $e^{\beta\mu}$, 
 cf.\ \eq\nr{n_actual}, 
 as this is needed in \eq\nr{comb}. 
 } 

Expanding the denominator of \eq\nr{n_canonical} in the fugacity
expansion and
identifying the contributions from the chosen sectors of the Fock space
in \eq\nr{n_exp}, we find
\ba
 p^{ }_1 & = & \frac{T }{V}   
 \Bigl[\, 
 \sum_{\vc{p}^{ }_\theta} 
   c^{ }_\theta\, 
   e^{- \beta E^{ }_{p_\theta} }
 + 
 \sum_{\vc{p}^{ }_\eta} 
   c^{ }_\eta\, 
   e^{- \beta E^{ }_{p_\eta} }
 \,\Bigr]
 \;, \la{p1_canonical} \\ 
 p^{ }_2 & = & \frac{T }{V}   
 \sum_{\vc{p}^{ }_\theta,\vc{p}^{ }_\eta}
 \Bigl[\, 
   c^{ }_{\theta,\eta}\,
   e^{- \beta E^{ }_{p_\theta,p_\eta} }
  - c^{ }_{\theta} c^{ }_{\eta} 
   e^{-\beta ( E^{ }_{p_\theta} + E^{ }_{p_\eta})} 
 \,\Bigr]
 \;. \la{p2_canonical}
\ea
For a perturbative evaluation, we write  
\be
 E^{ }_p \; \equiv \; 
 M^{ }_\rmi{rest} + \frac{p^2}{2 M^{ }_\rmi{kin}}
 \;,
\ee
where the Salpeter
correction (i.e.\ thermal shift of rest mass)
has been included in $M^{ }_\rmi{rest}$,
and $M^{ }_\rmi{kin}$ may similarly contain thermal effects. 
We note that, in the non-interacting limit, 
$
 E^{ }_{p_\theta,p_\eta}
 = 
 E^{ }_{p_\theta}
 + 
 E^{ }_{p_\eta} 
$
and
$ c^{ }_{\theta,\eta} = c^{ }_{\theta} c^{ }_{\eta} $.
Then $p^{ }_2 = 0$, whereas 
$n^{ }_\rmi{eq}$ from  
\eq\nr{neq_def} becomes (for $V\to\infty$)
\be
 n^{ }_\rmi{eq} 
 \; \approx \;
 2 \ds^{ } \Nc^{ } \int_\vc{p} e^{- \beta E^{ }_p}
 \;. \la{p1_LO} 
\ee

Proceeding to $\hat{p}^{ }_2$, 
we may follow the argument of 
ref.~\cite{binder} according to which
\eq\nr{p2_canonical} should be 
dominated by bound states at low temperatures. Let us write 
\be
 E^{ }_{p_\theta,p_\eta} = 
 2 M^{ }_\rmi{rest} + \frac{k^2}{4 M^{ }_\rmi{kin}} + E'
 \;, 
\ee
where $\vc{k} = \vc{p}^{ }_\theta + \vc{p}^{ }_\eta$ 
is the momentum of the center-of-mass motion, 
$k \equiv |\vc{k}|$,  and
$E'$ is the relative energy.
We may now write 
$
 \sum_{\vc{p}^{ }_\theta,\vc{p}^{ }_\eta} = 
 \sum_{\vc{k},E'}
$, 
and for the sum over $\vc{k}$ go over to infinite volume.
Furthermore, it is convenient to normalize $p^{ }_2$ as in 
\eq\nr{p2_hat}. Thereby
\be
  \hat{p}^{ }_2 
  \; = \;  
  \frac{\frac{1}{V} \sum_{\vc{p}^{ }_\theta,\vc{p}^{ }_\eta} 
 \bigl[
   c^{ }_{\theta,\eta}\,
   e^{- \beta E^{ }_{p_\theta,p_\eta} }
  - c^{ }_{\theta} c^{ }_{\eta} 
   e^{-\beta ( E^{ }_{p_\theta} + E^{ }_{p_\eta})}   
 \bigr]  
 }
 {
 \bigl[ 
  \frac{1}{V}  
  \bigl( 
   \sum_{\vc{p}^{ }_\theta} 
   c^{ }_\theta\, 
   e^{- \beta E^{ }_{p_\theta} }
   + 
   \sum_{\vc{p}^{ }_\eta} 
   c^{ }_\eta\, 
   e^{- \beta E^{ }_{p_\eta} }
  \bigr)
 \bigr]^2
 }
 \; \simeq \; 
 2 \biggl( \frac{\pi}{M^{ }_\rmi{kin} T} \biggr)^{3/2} 
 \sum_{- E' \gg T } \frac { 
   c^{ }_{\theta,\eta} }{ 
   c_\theta^2 
   } \,
   e^{- \beta E' }
 \;. \la{p2_canonical_2_appro}
\ee
Assuming that the contribution
of the $\ds^2(\Nc^2 - 1)$ octet degrees of freedom  
is exponentially suppressed, and omitting any hyperfine splitting, 
we can set $c^{ }_{\theta,\eta} \to \ds^2$. If we furthermore
assume that one bound state dominates, with the binding energy
by $\Delta E \simeq \alpha^2 M^{ }_\rmi{kin}/4$, and require 
a qualitatively correct limiting behaviour on the high-temperature side, 
we may set
\be
 T^3\, \hat{p}^{ }_2 
 \simeq 
 \frac{2}{\Nc^2} \biggl( \frac{\pi T}{M^{ }_\rmi{kin}} \biggr)^{3/2} 
  \Bigl(  e^{\beta \Delta E } - 1 \Bigr) 
 \;. \la{p2_canonical_3_appro}
\ee
We stress that this result should only be interpreted as 
an order-of-magnitude estimate, and that it is exponentially
sensitive to the choice of the value of $\alpha$ in $\Delta E$.

%
\section{Non-perturbative formulation}
\la{se:nonpert}

To go further, 
it is helpful to give an imaginary-time path-integral representation to the 
observables in \eqs\nr{n} and \nr{chi}. As we assume the fields $\theta,\eta$
to be non-relativistic, they propagate in one time direction only, and their
propagators are discontinuous across the imaginary-time interval. Therefore
some care is needed for defining a proper time ordering. 

For $n$, a convenient possibility is to split 
the time arguments by an infinitesimal amount, 
$
 n = \bigl\langle\, 
 \theta^\dagger(0,\vc{0})\, \theta(0^-,\vc{0}) + 
 \eta^\dagger(0,\vc{0})\, \eta(0^-,\vc{0}) 
 \,\bigr\rangle
$.
Antiperiodicity implies 
$\theta(0^-,\vc{0}) = -\theta(\beta,\vc{0})$, 
and we can subsequently use 
the Grassmann nature of the fields to 
anticommute $\theta(\beta,\vc{0})$ to the left. Therefore, 
\be
 n = \tr \bigl\langle\, 
 \theta(\beta,\vc{0}) \, \theta^\dagger(0,\vc{0})  + 
 \eta(\beta,\vc{0}) \, \eta^\dagger(0,\vc{0}) 
 \,\bigr\rangle
 \;. \la{n_tau}
\ee
We now denote (cf.\ appendix~A of ref.~\cite{swave})
\be
 \bigl\langle\,
  \theta(\beta,\vc{x})\, \theta^\dagger(0,\vc{x}) 
 \,\bigr\rangle^{ }_0
 \; \equiv \;
  e^{\beta\mu}\, G^{ }_\vc{x} 
 \;, \quad
 \bigl\langle\,
  \eta(\beta,\vc{x})\, \eta^\dagger(0,\vc{x}) 
 \,\bigr\rangle^{ }_0
  \; = \; 
  e^{\beta\mu}\, G^{*}_\vc{x}
 \;, 
\ee
where $\langle ... \rangle^{ }_0$ denotes a contraction of the 
Grassmann fields. Gauge fields are left to be 
averaged over later on, which is denoted by $\langle ... \rangle$. 
Then \eq\nr{n_tau} becomes
\be
 n = 2 e^{\beta\mu} \bigl\langle \re \tr G^{ }_\vc{0} \bigr\rangle
 \;. \la{n_latt}
\ee

For $\chi$, we point-split each $n$, and in addition  
make use of the fact that $\int_\vc{x} n(\tau,\vc{x})$ is a conserved
charge, whereby we can set the two $n$-operators at different times. So, 
\be
 \chi = \int_\vc{x} \Bigl\langle 
 \, \bigl [ 
    \theta^\dagger(\tau,\vc{x})\, \theta(\tau^-,\vc{x}) + 
    \eta^\dagger(\tau,\vc{x})\, \eta(\tau^-,\vc{x})
 \, \bigr]
 \, \bigl [ 
    \theta^\dagger(0,\vc{0})\, \theta(0^-,\vc{0}) + 
    \eta^\dagger(0,\vc{0})\, \eta(0^-,\vc{0})
 \, \bigr]
  - n^2 
 \Bigr\rangle 
 \;. \la{chi_tau}
\ee
Subsequently we can replace $\theta(0^-,\vc{0})$ through 
$-\theta(\beta,\vc{0})$, and again anticommute fields. 
This leads to 
\be
 \chi = \int_\vc{x}
 \Bigl\langle
   4 e^{2\beta\mu} \re\tr G^{ }_\vc{x} \re\tr G^{ }_\vc{0} + 
   2 e^{\beta\mu}  \re\tr G^{ }_\vc{0} - n^2  
 \Bigr\rangle 
 \;, \la{chi_nonpert}
\ee
where the middle term originates from contractions like 
\be
 \delta \chi = 
 \int_\vc{x} 
 \langle 
   \theta(\beta,\vc{0})\, \theta^\dagger(\tau,\vc{x})
 \rangle^{ }_0\,
 \langle 
   \theta(\tau,\vc{x})\, \theta^\dagger(0,\vc{0})
 \rangle^{ }_0
 \;, 
\ee
after making use of the semigroup property of the propagator.\footnote{%
 In a Dirac notation, this corresponds to the use of a completeness relation,
 $$
  \int_{\bf x } \langle \beta, {\bf 0} | \tau, {\bf x} \rangle
                \langle  \tau, {\bf x} | 0,    {\bf 0} \rangle
 = 
  \langle \beta, {\bf 0} | 0, {\bf 0} \rangle
 \;, \quad 0 < \tau < \beta
 \;.
 $$  
 } 

Now we can subtract $n$ of \eq\nr{n_latt} from $\chi$
of \eq\nr{chi_nonpert}
according to \eq\nr{p2}, thus obtaining a representation
for $p^{ }_2$. Moreover, normalizing according 
to \eq\nr{p2_hat}, where
$
 n^{ }_\rmi{eq} = 
 2 \langle \re \tr G^{ }_\vc{0}
 \rangle
$
according to \eqs\nr{neq_def} and \nr{n_latt}, we find  
\be
 \hat{p}^{ }_2 = 
 \frac{ 
 \int_\vc{x} 
 \bigl\{\, 
 \bigl\langle
  \re\tr G^{ }_\vc{x} \re\tr G^{ }_\vc{0} 
 \bigr\rangle
  - 
 \bigl\langle
  \re\tr G^{ }_\vc{0} 
 \bigr\rangle^2_{ }
 \,\bigr\} 
 }{
  2 
 \bigl\langle
  \re\tr G^{ }_\vc{0} 
 \bigr\rangle^2_{ }
 } 
 \;. \la{hat_p2_nonpert}
\ee
The numerator represents a ``disconnected'' contraction, with 
two heavy particle propagators not cancelling each other 
only because they are connected by gauge field lines. 

%
\section{Lattice measurement}
\la{se:latt}

In order to obtain non-perturbative information on the influence 
of bound states, we have measured 
$\hat{p}^{ }_2$ from \eq\nr{hat_p2_nonpert} 
with methods of non-relativistic lattice QCD. We have considered 
spinors with $s=\frac{1}{2}$, however we expect spin effects to 
be very small so that the results also apply to $s=0$.
For a good statistical precision, it is helpful to 
make use of translational invariance and 
rephrase the measurement of \eq\nr{hat_p2_nonpert} 
in analogy with \eq\nr{n}, 
\be
 T^3\, \hat{p}^{ }_2 = 
 \lim_{V\to\infty}
 \frac{T^3 V}{2} 
 \frac{ 
 \bigl\langle
   \mathcal{G}^2 
 \bigr\rangle
  - 
 \bigl\langle
  \mathcal{G} 
 \bigr\rangle^2_{ }
 }{
 \bigl\langle
  \mathcal{G} 
 \bigr\rangle^2_{ }
 } 
 \;, \quad
 \mathcal{G} \; \equiv \; 
 \frac{1}{V}
 \int_\vc{x} 
 \re\tr G^{ }_\vc{x}
 \;. \la{hat_p2_nonpert_alt}
\ee
The propagator $G^{ }_\vc{x}$ is constructed as explained
in ref.~\cite{swave}.\footnote{%
 For $M/T\to\infty$, 
 the numerator and denominator of  
 \eq\nr{hat_p2_nonpert_alt}
 correspond to the Polyakov loop
 susceptibility and expectation value squared, respectively.}

On a lattice, $T = 1/(N^{ }_\tau a^{ }_\tau)$ 
and $V = (N^{ }_s a^{ }_s)^3$, where
$a^{ }_\tau, a^{ }_s$ are the temporal and spatial lattice spacings 
and $N^{ }_\tau,N^{ }_s$ are the numbers of lattice points 
in these directions, respectively.
The details of the lattice setup
were summarized in ref.~\cite{pwave}; 
we have relied on refs.~\cite{lat0a,lat0b,lat1a,lat1b} for the adjustment of 
the bare parameters as well as for
the generation of the gauge configurations, both of which 
carry a substantial numerical cost.

\begin{figure}[t]

\hspace*{-0.1cm}
\centerline{%
 \epsfxsize=7.6cm\epsfbox{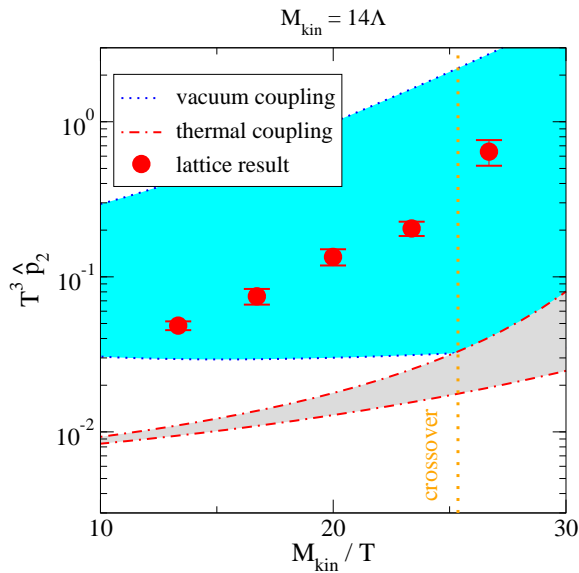}
}

\caption[a]{\small
 Comparison of an order-of-magnitude 
 estimate of $T^3 \hat{p}^{ }_2$
 from \se\ref{se:pert} 
 and a lattice estimate from  
 \se\ref{se:latt}. 
 The dashed line shows a temperature at which 
 confinement sets in.
 The errors of the lattice results
 are statistical only; systematic uncertainties 
 could be as large as $\sim 50\%$.  
 In any case, 
 based on this test, 
 a vacuum-like coupling 
 performs best at low temperatures, whereas
 towards high temperatures the slope seen in the  
 data agrees better with a thermal 
 scale choice (i.e.\ $\bmu \sim \pi T$).
}

\la{fig:hatp2}
\end{figure}

The key idea of the lattice test is that the importance of 
any effects associated with bound states depends on the ratio
$ \Delta E /T \sim \alpha^2 M/T$, where $\Delta E$ is the 
binding energy 
and $M$ is the dark matter mass scale. In the following, 
we denote by $\Lambda$ the $\msbar$ scale parameter.
In cosmological applications, the phenomenologically relevant
mass scale is $M \;\gsim\; 1$~TeV $\gg 10^3_{ }\Lambda$, 
and correspondingly the coupling $\alpha\sim 0.1$ is ``small''. In this 
situation bound-state effects are expected to be large only in the regime
$M/T \;\gsim\; 1/\alpha^2\sim 100$. 
In contrast, lattice simulations are best suited
to moderate temperatures, $T/\Lambda \sim 0.5 ... 1.0$, and 
a situation without large scale hierarchies.
Then the coupling
is ``large'', $\alpha \gsim 0.3$, and bound-state effects are 
important already for $M/T \sim 10...30$. 
The idea is now that if we can use lattice 
to scrutinize analytic
estimates in the domain of large couplings, we should be confident
that they apply in the cosmological domain of small couplings.  

The results of the lattice measurements
are shown in \fig\ref{fig:hatp2}, where they are also  
compared with the order-of-magnitude estimate from 
\se\ref{se:pert}. The dominant uncertainty of the latter 
is the choice of $\alpha$. By a ``vacuum coupling'' 
we indicate that $\alpha$ has been evaluated 
at a scale $(0.5 ... 2.0) e^{-\gammaE}/a$, 
where $a = 2/(M \alpha)$ is the Bohr radius; 
the factor $e^{-\gammaE}$ is inspired by 
refs.~\cite{pot1,pot4}; and we have solved 
the implicit equation for $\alpha$ numerically,  
by employing 2-loop running.
By a ``thermal coupling'' we indicate the dimensionally reduced value,
as specified in appendix~A of ref.~\cite{stop}. 

In view of the experience from 
ref.~\cite{pwave}, where other observables were measured
in the same temperature range, as well as the exponential
dependence on $\alpha$,
the rough qualitative agreement between the lattice
and analytic results seen in \fig\ref{fig:hatp2}
should be considered reasonable. 
The lesson we draw is that at low temperatures
the vacuum coupling should be a fair choice, whereas at high 
temperatures, where bound states are less prominent
and ultimately dissolve, the results
tend gradually towards a thermal value
(though they do not reach it within the domain of large $\alpha$).  
In \se\ref{se:dm} we interpolate between these two possibilities. 

%
\section{A dark matter application}
\la{se:dm}

Having tested $\hat{p}^{ }_2$
from \eq\nr{p2_canonical_3_appro} against lattice data in \se\ref{se:latt}, 
we are now ready to apply the same estimate to a simple but realistic 
cosmological computation. In this case we add a neutral field to the 
model (as dark matter proper), 
and let the charged field (mediator) be in general heavier, 
by an amount $\Delta M$. 

Specifically, we consider the setup reviewed in 
refs.~\cite{giv,mg3} and recently studied for bound-state effects
in refs.~\cite{ll,mrss,klz,stop,hp,sb,fls}, 
in which the dark matter particle is a singlet
Majorana fermion, and the dark sector also contains 
a strongly coupled scalar mediator, a ``stop''. 
If the stop is not much heavier than the Majorana fermion, 
strong interactions between a stop and antistop open up a 
very efficient annihilation channel in the early universe, 
reducing the dark matter abundance to an acceptable level even in 
the multi-TeV mass range. 
Simultaneously, the $p$-wave suppressed annihilations of the Majorana
fermion at low energies guarantee that constraints from indirect detection
can be satisfied. The direct detection constraints are weak, 
if the Yukawa interaction couples the stop dominantly to 
3rd generation quarks~\cite{sb}. Furthermore collider constraints can 
be evaded if the stops are heavier than $\sim 1$~TeV. 

The same model was studied within the current formalism 
in refs.~\cite{stop,sb}, however under
the assumption $\hat{p}^{ }_2 = 0$, whereby
$ e^{\beta \mu}\, n^{ }_\rmi{eq} = n$ according to \eq\nr{n_actual}. 
This led to the problem that 
bound-state effects became extremely large 
at $z \; \equiv \; M/T \gg 10^3$. 
In order to avoid this problem, 
the mass splitting $\Delta M$ was chosen large enough to satisfy
$2\Delta M > \Delta E$, so that bound states
of stops were always heavier than scattering states 
of Majorana fermions, and thus ultimately exponentially suppressed.
The equations were integrated down to $z = 10^3$.

\begin{figure}[t]

\hspace*{-0.1cm}
\centerline{%
 \epsfxsize=7.6cm\epsfbox{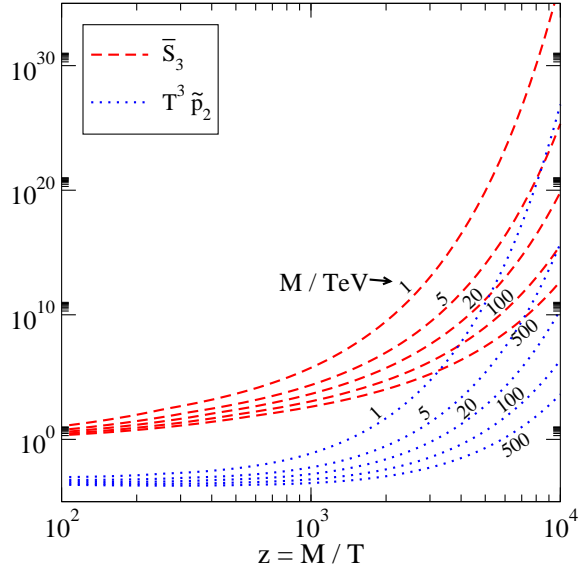}
}

\caption[a]{\small
  Values of the thermally averaged Sommerfeld factor
  $\bar{S}^{ }_3$ 
  (cf.\ \eq\nr{Sbar_order_of_mag})
  and the rescaled susceptibility $T^3\, {\tilde p}^{ }_2$ 
  (cf.\ \eq\nr{hatp2_order_of_mag})
  for $M = 1...500$~TeV.
}

\la{fig:hatp2_barS3}
\end{figure}

We have now included $\hat{p}^{ }_2$ in the dynamics 
described by \eq\nr{comb}, 
by solving for $e^{\beta\mu}\, n^{ }_\rmi{eq}$ from \eq\nr{n_actual}. 
The presence of the neutral field implies that
\be
 n^{ }_\rmi{eq} \simeq 2 
 \biggl( \frac{MT}{2\pi} \biggr)^{\fr32}
 e^{-\beta M}
 \, 
 \Bigl(
  1 + \Nc^{ } \, e^{-\beta \Delta M^{ }_\T} 
 \Bigr)
 \;, \la{neq_cosmo}
\ee
where the thermally modified mass difference 
$
 \Delta M^{ }_\T
$
is given in \eq(4.8) of ref.~\cite{stop}, and we have dropped
the subscript from $M^{ }_\rmi{kin}$ for simplicity. Given the normalization
by $n_\rmi{eq}^2$ (cf.\ \eq\nr{p2_hat}), the order-of-magnitude
estimate from \eq\nr{p2_canonical_3_appro} becomes
\be
 \hat{p}^{ }_2 
 \; \simeq \; 
 \frac{\Nc^2}
 {\bigl(\Nc^{ } + e^{\beta\Delta M^{ }_{\T} } \bigr)^2}
 \, {\tilde p}^{ }_2
 \;, \quad
 T^3\, {\tilde p}^{ }_2
 \; \equiv \; 
  \frac{2}{\Nc^2} 
  \biggl( \frac{\pi T}{M} \biggr)^{3/2} 
  \Bigl(  e^{\beta \Delta E } - 1 \Bigr) 
 \;, \la{hatp2_order_of_mag}
\ee
where $\Delta E = \alpha^2 M/4$, and we have for convenience
defined a $\Delta M^{ }_\T$-independent $\tilde{p}^{ }_2$.
The corresponding approximation 
for the attractive Sommerfeld factor reads~\cite{stop}
\be
 \bar{S}^{ }_3 \;\approx\;
 \biggl( \frac{4\pi}{MT} \biggr)^{\fr32} 
 \frac{ e^{\beta \Delta E } }{\pi a^3}
 \;, \la{Sbar_order_of_mag}
\ee
where $a = 2/(M\alpha)$ is the Bohr radius. 
Inspired by the tests in \se\ref{se:latt}, 
at low temperatures
the coupling $\alpha$ is evaluated at the $\msbar$ 
scale $\sim e^{-\gammaE}/a$, 
and at high temperatures we use a thermal coupling; 
the crossover takes place at $z \approx 250 ... 600$
for $M = 1 ... 500$~TeV.
This $\bar{S}^{ }_3$ attaches rather smoothly 
to the more elaborate results described in ref.~\cite{stop}; 
in practice we can use the simplified expression from 
\eq\nr{Sbar_order_of_mag}
at $z \gsim 200$. 
The repulsive Sommerfeld factors $\bar{S}^{ }_{4,5}$~\cite{stop},
which are not important at late times,  
are frozen to their values at $z\simeq 200$.

\begin{figure}[t]

\hspace*{-0.1cm}
\centerline{%
 \epsfxsize=7.6cm\epsfbox{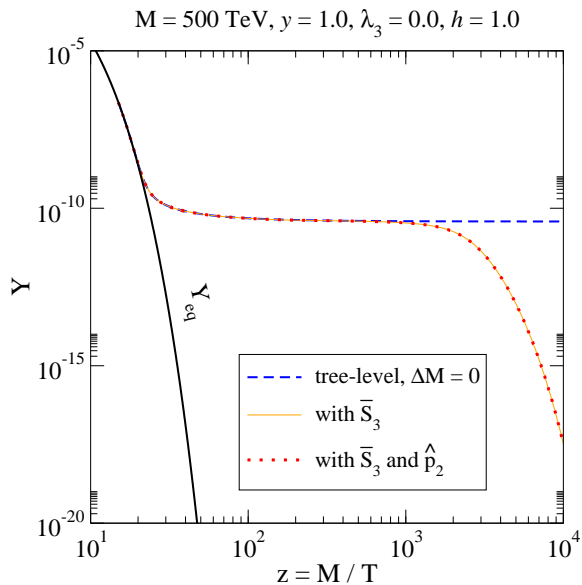}
}

\caption[a]{\small
 Examples of a solution, for $M = 500$~TeV, 
 with tree-level annihilation rates (``tree-level'')
 and after including a thermally averaged Sommerfeld factor 
 (``with $\bar{S}^{ }_3$'') and
 a susceptibility
 (``with $\bar{S}^{ }_3$ and $\hat{p}^{ }_2$''). 
 The symbols $y, \lambda^{ }_3,h$ refer to couplings
 defined in ref.~\cite{stop}, whose precise values have
 little impact on the general pattern. This plot assumes
 that kinetic/ionization equilibrium is maintained in
 the dark sector. 
}

\la{fig:Y_M500000_hatp2}
\end{figure}

Numerical values of $T^3 {\tilde p}^{ }_2$ and $\bar{S}^{ }_3$
are shown in \fig\ref{fig:hatp2_barS3}. 
As anticipated in ref.~\cite{binder}, the presence 
of a non-zero $\hat{p}^{ }_2$ in principle ``regulates'' 
the late-time behaviour of the system: 
the growth of $\langle\, \sigma v \,\rangle$, which is
proportional to $\bar{S}^{ }_3$, is compensated for by 
the growth of $\hat{p}^{ }_2$, which appears in the 
numerator (cf.\ \eqs\nr{comb} and \nr{n_actual}). 
Alas, we find that in practice this regulation is {\em not efficient}
in this model. This can be understood by inspecting the combination
\be
 8 \hat{p}^{ }_2 n = 8 \, T^3 \hat{p}^{ }_2
 \, \Bigl( \frac{s}{T^3} \Bigr) \, Y 
 \;, \la{8p2n}
\ee
that appears in \eq\nr{n_actual} (here $Y \equiv n/s$). 
The entropy density is $s/T^3 \lsim 50$. 
Recalling 
\be
   \Omega^{ }_\rmi{dm} h^2 \approx
           \frac{Y(z^{ }_\rmi{final})\,M}
           {[3.645 \times 10^{-12}\,\mbox{TeV}]} \approx 0.12
 \;,  
\ee
we are interested in yields 
$Y \simeq 10^{-13}$. 
According to \eq\nr{8p2n},
we would then need $T^3 \hat{p}^{ }_2\gg 10^{11}$, 
in order to have $8 \hat{p}^{ }_2 n \gg 1$  
and thus a substantial regularization through $\hat{p}^{ }_2$. 

Now, according to \fig\ref{fig:hatp2_barS3}, 
values $T^3 \hat{p}^{ }_2\gsim 10^{11}$
can indeed be found if $M \lsim 20$~TeV, 
however they only set in at large $z$. Unfortunately, 
by this time $\bar{S}^{ }_3 \gg 10^{10}$, 
whereby $Y \simeq 10^{-13}$ can actually not be found in this model. 
The situation is illustrated in 
\fig\ref{fig:Y_M500000_hatp2} for an extreme case $M = 500$~TeV, 
chosen to push the initial $Y$ as large as possible. It is clear that 
the large $\bar{S}^{ }_3$ rapidly pulls $Y$ down to such small 
values that the increasing $\hat{p}^{ }_2$ has no visible effect. 
To summarize, we find no stabilizing effect from $\hat{p}^{ }_2$ in the
whole mass range considered ($1 ... 500$~TeV), 
leaving $\Delta M > \Delta E/2$ as the
only possible (equilibrium) regulator. 

%
\section{Conclusions}
\la{se:concl}

The purpose of this paper has been to explore the implications of the modified 
Lee-Weinberg equation (cf.\ \eq\nr{comb})
put forward in ref.~\cite{binder}. On one hand, 
we have shown how the coefficient $\hat{p}^{ }_2$, 
which captures the essence of the Saha equation (cf.\ \eq\nr{n_actual}), 
can be related to a ``susceptibility'' (cf.\ \eq\nr{hat_p2_nonpert_alt}), 
which can be measured non-perturbatively within a non-relativistic 
lattice QCD framework (cf.\ \fig\ref{fig:hatp2}). 
On the other hand, we have shown how $\hat{p}^{ }_2$
can be used in a practical dark matter computation, where
it implements ionization equilibrium in accordance with 
the Saha equation, and therefore guarantees that bound states
appear with their thermal abundance
(this assumption ceases to be valid at very low temperatures). 
As proposed in ref.~\cite{binder}, the presence of
$\hat{p}^{ }_2$ can in principle
regulate the late-time behaviour of the system, 
in addition to the regularization provided
by an explicit mass difference $\Delta M$ in the dark sector,
or the non-equilibrium effects that inevitably
take over at the very end.  

However, considering a concrete model with a strongly interacting
mediator, we find that in practice the regularization by $\hat{p}^{ }_2$
is insufficient to make the system viable if $2 \Delta M < \Delta E$, 
where $\Delta E$ is the binding energy for bound states in the mediator
sector (cf.\ \fig\ref{fig:Y_M500000_hatp2} and the discussion
around \eq\nr{8p2n}). This implies that 
the viable domain
remains sensitive to the value of $\Delta M$ in this model
(nevertheless the viable domain extends at least  
to the multi-TeV range as discussed in refs.~\cite{stop,sb}). 
Whether other models could behave differently is not clear
at the moment, even if we note that in general
$\bar{S}^{ }_3 / (T^3 \hat{p}^{ }_2) \simeq (M \alpha / T)^3 \gg 1$
at low temperatures, suggesting that $\hat{p}^{ }_2$ is not
sufficient to compensate for the effect of $\bar{S}^{ }_3$.

It is perhaps prudent to stress that 
our current analytic values of $\hat{p}^{ }_2$ amount just to an 
order-of-magnitude estime, originating from 
a Coulomb-like ground-state binding energy. At least  
on the high-temperature side, this could in principle be promoted 
into a consistent leading-order perturbative computation, 
however this is demanding, given that 
\eq\nr{hat_p2_nonpert} originates from a disconnected contraction, 
and is therefore of 3-loop order, i.e.\ $\rmO(\alpha^2)$. 

%
\section*{Acknowledgements}

We thank the FASTSUM collaboration for providing the unquenched 
gauge configurations used in our lattice measurements. 
S.B.\ thanks AEC/ITP of the University of Bern for hospitality
during initial stages of this work. 
S.K.\ was
supported by the National Research Foundation of Korea under grant
No.\ 2018R1A2A2A05018231 funded by the Korean government (MEST) and in
part by NRF-2008-000458. M.L.\ was supported 
by the Swiss National Science Foundation
(SNF) under grant 200020-168988.

\small{
%

}


\begin{thebibliography}{99}

\bibitem{clas1}
  B.W.~Lee and S.~Weinberg, 
  {\it Cosmological Lower Bound on Heavy Neutrino Masses}, 
  Phys.\ Rev.\ Lett.\  {39} (1977) 165.

\bibitem{clas2}
  J.~Bernstein, L.S.~Brown and G.~Feinberg,
  {\it The Cosmological Heavy Neutrino Problem Revisited,}
  Phys.\ Rev.\ D {32} (1985) 3261.

\bibitem{old1}
  K.~Griest and D.~Seckel,
  {\it Three exceptions in the calculation of relic abundances,}
  Phys.\ Rev.\ D {43} (1991) 3191.

\bibitem{old32}
  W.~Detmold, M.~McCullough and A.~Pochinsky,
  {\it Dark Nuclei I: Cosmology and Indirect Detection,}
  Phys.\ Rev.\ D {90} (2014) 115013
  [1406.2276].

\bibitem{old4}
  B.~von Harling and K.~Petraki,
  {\it Bound-state formation for thermal relic dark matter and unitarity,}
  JCAP {12} (2014) 033
  [1407.7874].

\bibitem{seq}
  F.~Karsch, D.~Kharzeev and H.~Satz,
  {\it Sequential charmonium dissociation,}
  Phys.\ Lett.\ B {637} (2006) 75
  [hep-ph/0512239].

\bibitem{chemical}
  D.~B\"odeker and M.~Laine,
  {\it Heavy quark chemical equilibration rate as a transport coefficient,}
  JHEP {07} (2012) 130
  [1205.4987].

\bibitem{bodwin}
  G.T.~Bodwin, E.~Braaten and G.P.~Lepage,
  {\it Rigorous QCD analysis of inclusive annihilation and 
  production of heavy quarkonium,}
  Phys.\ Rev.\ D {51} (1995) 1125; 
  {\it ibid.} {55} (1997) 5853 (E)
  [hep-ph/9407339].

\bibitem{swave}
  S.~Kim and M.~Laine,
  {\it Rapid thermal co-annihilation through bound states in QCD,}
  JHEP {07} (2016) 143
  [1602.08105].

\bibitem{binder}
  T.~Binder, L.~Covi and K.~Mukaida,
  {\it Dark Matter Sommerfeld-enhanced annihilation and Bound-state 
  decay at finite temperature,}
  Phys.\ Rev.\ D {98} (2018) 115023
  [1808.06472].

\bibitem{kubo}
  D.~B\"odeker and M.~Laine,
  {\it Kubo relations and radiative corrections for lepton number washout,}
  JCAP {05} (2014) 041
  [1403.2755].

\bibitem{sangel}
  D.~B\"odeker and M.~Sangel,
  {\it Order $g^2$ susceptibilities in the symmetric phase
  of the Standard Model,}
  JCAP {04} (2015) 040
  [1501.03151].

\bibitem{pwave}
  S.~Kim and M.~Laine,
  {\it Studies of a thermally averaged $p$-wave Sommerfeld factor,}
  Phys.\ Lett.\ B {795} (2019) 469
  [1904.07882].

\bibitem{lat0a}
  R.G.~Edwards, B.~Joo and H.W.~Lin,
  {\it Tuning for Three-flavors of Anisotropic Clover Fermions
  with Stout-link Smearing,}
  Phys.\ Rev.\ D {78} (2008) 054501
  [0803.3960].

\bibitem{lat0b}
  H.W.~Lin {\it et al.} [Hadron Spectrum Collaboration],
  {\it First results from 2+1 dynamical quark flavors on an anisotropic
  lattice: Light-hadron spectroscopy and setting the strange-quark mass,}
  Phys.\ Rev.\ D {79} (2009) 034502
  [0810.3588].

\bibitem{lat1a}
  C.~Allton {\it et al.},
  {\it 2+1 flavour thermal studies on an anisotropic lattice,}
  PoS LATTICE {2013} (2014) 151
  [1401.2116].

\bibitem{lat1b}
  G.~Aarts {\it et al}, 
  {\it The bottomonium spectrum at finite temperature 
  from N$_{f}$ = 2 + 1 lattice QCD,}
  JHEP {07} (2014) 097
  [1402.6210].

\bibitem{pot1}
  Y.~Schr\"oder,
  {\it The Static potential in QCD to two loops,}
  Phys.\ Lett.\ B {447} (1999) 321
  [hep-ph/9812205].

\bibitem{pot4}
  R.N.~Lee, A.V.~Smirnov, V.A.~Smirnov and M.~Steinhauser,
  {\it Analytic three-loop static potential,}
  Phys.\ Rev.\ D {94} (2016) 054029
  [1608.02603].

\bibitem{giv}
  M.~Garny, A.~Ibarra and S.~Vogl,
  {\it Signatures of Majorana dark matter with $t$-channel mediators,}
  Int.\ J.\ Mod.\ Phys.\ D {24} (2015) 1530019
  [1503.01500].

\bibitem{mg3}
  M.~Garny, J.~Heisig, M.~Hufnagel and B.~L\"ulf,
  {\it Top-philic dark matter within and beyond the WIMP paradigm,}
  Phys.\ Rev.\ D {97} (2018) 075002
  [1802.00814].

\bibitem{ll}
  S.P.~Liew and F.~Luo,
  {\it Effects of QCD bound states on dark matter relic abundance,}
  JHEP {02} (2017) 091
  [1611.08133].

\bibitem{mrss}
  A.~Mitridate, M.~Redi, J.~Smirnov and A.~Strumia,
  {\it Cosmological Implications of Dark Matter Bound States,}
  JCAP {05} (2017) 006
  [1702.01141].

\bibitem{klz}
  W.Y.~Keung, I.~Low and Y.~Zhang,
  {\it A Reappraisal on Dark Matter Co-annihilating with a Top/Bottom
  Partner,}
  Phys.\ Rev.\ D {96} (2017) 015008
  [1703.02977].

\bibitem{stop}
  S.~Biondini and M.~Laine,
  {\it Thermal dark matter co-annihilating with 
  a strongly interacting scalar,}
  JHEP {04} (2018) 072
  [1801.05821].

\bibitem{hp}
  J.~Harz and K.~Petraki,
  {\it Radiative bound-state formation in unbroken perturbative 
  non-Abelian theories and implications for dark matter,}
  JHEP {07} (2018) 096
  [1805.01200].

\bibitem{sb}
  S.~Biondini and S.~Vogl,
  {\it Coloured coannihilations: 
  Dark matter phenomenology meets non-relativistic EFTs,}
  JHEP {02} (2019) 016
  [1811.02581].

\bibitem{fls}
  H.~Fukuda, F.~Luo and S.~Shirai,
  {\it How Heavy can Neutralino Dark Matter be?,}
  JHEP {04} (2019) 107
  [1812.02066].
  
\end{thebibliography}
\end{document}